\documentclass[runningheads]{llncs}

\usepackage{ppack}

\begin{document}
\title{Differential experiments using parallel alternative operations
}
%
%
\author{Marco Calderini\inst{1} \and
Roberto Civino\inst{2} \and
Riccardo Invernizzi\inst{3}}
\authorrunning{M. Calderini et al.}
%
\institute{University of Trento, \email{marco.calderini@unitn.it} \and
University of L'Aquila, 
\email{roberto.civino@univaq.it} \and
KU Leuven, \email{riccardo.invernizzi@kuleuven.be}}
\maketitle              
\begin{abstract}
The use of alternative operations in differential cryptanalysis, or alternative notions of differentials, are lately receiving increasing attention. Recently, Civino et al.~managed to design a block cipher which is secure w.r.t.\ classical differential cryptanalysis performed using XOR-differentials, but weaker with respect to the attack based on an alternative difference operation acting on the first s-box of the block. We extend this result to parallel alternative operations, i.e.\ acting on each s-box of the block. First, we recall the mathematical framework needed to define and use such operations. After that, we perform some differential experiments against a toy cipher and compare the effectiveness of the attack w.r.t.\ the one that uses XOR-differentials.

\keywords{differential cryptanalysis; alternative operations; distinguisher; block ciphers.\\[0,5cm]
}
\end{abstract}

\section{Introduction}

Differential cryptanalysis is a powerful tool introduced in the beginning of the 90's to attack some cryptographic symmetric primitives, namely block ciphers~\cite{biham1991differential}.
The attack, which has later been generalised \cite{BBS99,K94,W99},  is tipically a chosen-plaintext attack that takes advantage of non-uniform relations between input differences and corresponding output differences.

To mitigate vulnerability to these attack methods, the cryptographic transformations employed within the substitution boxes (s-boxes) of the cipher should aim for the lowest possible level of differential uniformity~\cite{N93} (for a comprehensive exploration of the differential uniformity of vectorial Boolean functions, readers can refer to Mesnager et al.'s survey~\cite{MMM22}). It is essential to emphasize that the calculation of differential uniformity is based on the XOR operation. Indeed, in a traditional scenario of cryptanalysis of block ciphers, the difference operation classically taken into consideration by both designers and cryptanalysts is the one used to mix the key during the encryption process. In many cases, this operation is the bit-wise addition modulo two, i.e., the XOR. Nevertheless, it is worth noting that alternative types of operations may also be contemplated.
For example, Berson introduces the modular difference to study the MD/SHA family of hash functions~\cite{B92}, and similar method has been used~\cite{AS11} to cryptanalyze the block cipher PRESENT~\cite{bogdanov2007present}.
Borisov et al.~\cite{BCJW02} proposed a new type of differential known as multiplicative differential to attack IDEA \cite{IDEA}. This inspired the definition of c-differential uniformity~\cite{CDIFF}, which has been extensively studied in the last couple of years, even if the cryptographic implication of such $c$-differential uniformity on attacking block ciphers remains a subject of ongoing debate~\cite{DANIELE}.
In 2019, Civino et al.\ showed that a differential attack making use of alternative differences may be effective against XOR-based ciphers that are resistant to the classical differential attack~\cite{civino}. More precisely, they designed a small-scale substitution-permutation network (SPN) inspired to the block cipher PRESENT, with 5 s-boxes of 3 bits each.  They introduced a new sum on the whole message space that acts as the XOR on the last 4 s-boxes, while on the first one matches with one of the alternative sums defined by Calderini et al.~\cite{calderini}, coming from another elementary abelian regular group of translations (\emph{translation groups} in short). Using such operation, they were able to mount a distinguishing attack on 5 rounds of the cipher. Moreover, they showed that this result cannot be obtained with the traditional differential approach, i.e., looking at the distribution of classical differentials. 

In this work we show that this idea can be extended to the whole block, attacking all the s-boxes at the same time. The difference operator that we consider comes again from the family of alternative operations introduced by Calderini et al.~\cite{calderini}.
Although it could be made more general without efforts, for the ease of description we focus here on the case of 
translation-based ciphers~\cite{caranti2009some}, which are the most common form of SPNs nowadays, where the encryption is realised by  subsequent iterations of a non-linear s-box layer, a (usually) linear permutation and a XOR-based key addition. \\
 In order to guarantee deterministic propagation of differences through the diffusion layers of the ciphers, as happens in the classical scenario, we  need to characterise the XOR-linear bijective maps that are also linear with respect to another (parallel) sum. As we will discuss later, the mentioned problem is of general interest in cryptography, and results in this direction will produce examples of XOR-based trapdoor ciphers for which a non-XOR distinguisher may exist. Unfortunately, only partial solutions to this problem are known~\cite{civino,brunetta2019hidden,aragona2019regular}.
Keeping our focus in this direction,
after performing some computational experiments, we are able to design a toy cipher similar to the one in Civino et al.~\cite{civino}, with 4 parallel s-boxes of 4 bits each\footnote{It is important to specify that the design technique is completely scalable and that we decided to perform experiments on a small-sized cipher only for a matter of efficiency.} together with an alternative parallel operation $\circ$ that can be used to attack it. 
A computer-aided direct check shows that the diffusion layer of the proposed cipher is a linear permutation w.r.t.\ both the operations $+$ and $\circ$, which is the pivotal condition for the success of the attack.\\
We  show the results of a distinguishing attack for a number of rounds up to 17, concluding that differentials based on our alternative operation have much higher probabilities. Moreover, the difference in probability between alternative differentials and classical differentials that we obtain is higher than the one in Civino et al.~\cite{civino}, showing the effectiveness of our approach.

The paper is organised as follows. In Section \ref{sec2} we introduce the notation and describe the general setting for our attack. A brief summary on the construction of alternative operations coming from elementary abelian regular groups is given in Section~\ref{sec3}. In Section~\ref{sec4} we design a 16-bit cipher and a suitable parallel alternative operation, and perform experiments to study its resistance to differential cryptanalysis. We show the consistent improvement of our approach with respect to the classical one. We conclude the paper with the discussion of some open problems in Section~\ref{sec_fin}. 


\section{Preliminaries}
\label{sec2}
Let $V = (\F_2)^n$ be a binary vector space, with canonical basis $e_1,\dots,e_n$, which will represent the plaintext-ciphertext space of an $n$-bit block cipher. We denote by
\[ T := \{\sg_a \mid a\in V, \sg_a:x \mapsto x+a \} < \sym(V)\]
the group of translations on $V$. We stress that the action of this translation group on the message space $V$ represents the XOR-based key-addition layer in a block cipher. Let us also notice that $T$ as a subgroup of the symmetric group $\sym(V)$ is elementary, abelian and regular. We recall here the definition of regularity.
\begin{definition}
    A permutation group $G$ acting transitively on a set $V$ is said to be regular if for all $v\in V$ the stabilizer of $G$ at $v$ $G_v \defeq \{g \in G \mid vg=v\}$  is trivial.
\end{definition}
It is well known that any other elementary abelian regular subgroup of $\sym(V)$ is conjugated to $T$.
\begin{theorem}[\cite{dixon1971maximal}]
    \label{teo:coniug_general}
    Let $\mathcal{T} < \sym(V)$ be an elementary abelian regular subgroup. Then, there exists $g \in \sym(V)$ such that $\mathcal{T} = \T^g = g^{-1}\T g$.
\end{theorem}
Let us now show how to define another operation on $V$ starting from another elementary abelian regular group of translations.
\begin{definition}
    \label{def:circ}
    Let $\mathcal{T}<\sym(V)$ be an elementary abelian regular group. Let us define an additive group operation $\circ$ on $V$ by letting for each $a,b$ in $V$
    \[ a \circ b \defeq a \tau_b,\]
    where $\tau_b$ is the unique element of $\mathcal{T}$ sending $0$ to $b$.
\end{definition}

\begin{proposition}
   If $\circ$ is defined as above, then  $(V, \circ)$ is a vector space over $\F_2$, with associated translation group $\tto = \tcal$. Moreover, $(V, \circ) \cong (V, +)$. 
\end{proposition}
The subspaces introduced below are essential to understand the structure of alternative operations coming from translation groups. We refer to Civino et al.\ for a more detailed discussion~\cite{civino}.
\begin{definition}\label{def}
    Given an operation $\circ$ as above, a vector $k \in V$ is called a \emph{weak key} if for each $x \in V$ it holds $x+k = x \circ k$. The set
    $$ \wo \defeq \{k \mid k \in V, \ k \text{ is a weak key} \}$$
    is called the \emph{weak-key space}, and is a subspace of both  $(V, +)$ and $(V, \circ)$. We denote by $d\geq 1$ its dimension.
    Moreover, let us define a dot product on $V$ letting for each $a,b\in V$  \[a \cdot b \defeq a + b + a \circ b.\]
    The set of elements that can be expressed as dot products is denoted by
    $$\uo \defeq \{x \cdot y \mid x,y \in V \}$$
    and is called the \emph{set of errors}. \\
    Finally, denoting by $\gl(V, +)$ and $\gl(V, \circ)$ the groups of linear permutations with respect $+$ and $\circ$ respectively, we define
    $$ \ho := \gl(V, +) \cap \gl(V, \circ).$$
\end{definition}

We now briefly present the impact of an alternative sum $\circ$ on the differential cryptanalysis of SPNs. The classical differential attack relies on the property that each  XOR-difference is maintained the same after the key is XOR-ed to the state. This is not the case when considering $\circ$-differences. Indeed, let us consider two inputs with difference $\Dt$, denoted by $x$ and $x \circ \Dt$. After the key addition, the difference becomes
$$ (x+k)\circ ((x\circ \Dt)+k)\eqdef \Dt^\circ.$$
However, it can be shown~\cite{civino} that if $T < \agl(V,\circ)$ and $T_\circ < \agl(V,+)$, then
\begin{equation}\label{eq}
 \Dt^\circ = \Dt + k\cdot \Dt,
 \end{equation}
i.e., in a particular setting, the output difference after the key-addition layer can be expressed in terms of the dot product introduced in Definition~\ref{def}.
By definition, $k\cdot \Dt$ belongs to $\uo$, therefore the number of possible output differences is bounded by $|\uo|$. The presence of the \emph{error} in Eq.~\eqref{eq}, of course, forces us to consider $\circ$-differential probabilities introduced by the key-addition layer, unlike in the classical case, yielding a disadvantage in terms of the final probability of the differential propagation.\\
On the other hand though, the  s-box is usually designed to have the lowest possible differential uniformity w.r.t.\ the XOR. This may no longer be  true w.r.t.\ the operation $\circ$. A higher differential uniformity creates trails with higher probability for $\circ$ and counterbalances the effect of  differential probability introduced by the key addition. \\
Finally, and more importantly, the diffusion layer $\lambda$ of an SPN is usually a XOR-linear map. In order to mount a successful $\circ$-differential attack we need $\lambda$ to be $\circ$-linear as well, i.e., $\lambda \in \ho$. Otherwise, $\lambda$ would be a $\circ$-non-linear map and the effect of block-sized differential probabilities introduced by the diffusion layer would make the approach completely ineffective.  This represents a strong motivation to the study of $\ho$.\\

We are now ready to show explicitly how operations coming from new translation groups are constructed.

\subsection{Construction of alternative operations}
\label{sec3}
For the reason explained above (see Eq.~\eqref{eq}), it is convenient to consider operations $\circ$ on $V$ coming from a translation group $T < \sym(V)$ such that $T < \agl(V,\circ)$ and $T_\circ < \agl(V,+)$, which is the setting in which we will assume to be from now on. We will make use of the construction of such operations as presented in Calderini et al.~\cite{calderini}, but we will omit here many of the details, that the interested reader can find in the cited paper. 

Recall that we denote $n = \dim(V)$ and that we have $1\leq d=\dim(\wo) \leq n-2$~\cite{calderini}. We will focus on the particular case $d = n-2$. The reason for this is that the case when the dimension of the weak-key space reaches its upper bound is one of those in which the structure of $H_\circ$ is known (see the following Theorem~\ref{teo:caratt_ho}). Thanks to Calderini et al.~\cite[Theorem 3.9]{calderini}, we may assume, up to conjugation, that $W_\circ$ is spanned by $\{e_3,\dots,e_n\}$. In this setting, from Calderini et al.~\cite[Theorem 3.11]{calderini} (but see also Civino et al.~\cite[Theorem 3.3]{civino}) we have 
\[a \circ e_i = a \tau_{e_i} = aM_{e_i} + e_i,
\] where
\[ 
M_{e_1} = 
\left(\begin{array}{@{}c|c@{}}
    \ind{2} &
    \begin{matrix}
        \textbf{0} \\
        \b
    \end{matrix}
    \\
    \hline
    \mathbb{0}_{n-2, 2} &
    \ind{n-2}
\end{array}\right),
\ M_{e_2} = 
\left(\begin{array}{@{}c|c@{}}
    \ind{2} &
    \begin{matrix}
        \b \\
        \textbf{0}
    \end{matrix}
    \\
    \hline
    \mathbb{0}_{n-2, 2} &
    \ind{n-2}
\end{array}\right),
\]
and $M_{e_j} = \ind{n}$ for $j \geq 3$, where $\ind{k}$ denotes the the identity matrix of size $k\times k$ and  $\mathbb{0}_{k, \ell}$ is the zero matrix of size $k\times \ell$. The element $\b$ is a non-zero vector in $(\F_2)^{n-2}$, which completely determines $\circ$. Once the operation is defined of the basis, it is easy to compute $a \circ b$, for $a,b \in V$. \\

Let $r$ and $s$ be two positive integers, we will denote by $(\F_2)^{r\times s}$ the set of matrices of dimension $r\times s$. The following result is due to Civino et al.~\cite[Theorem 5.3]{civino} and characterises $H_\circ$ in the case $d=n-2$.

\begin{theorem}
    \label{teo:caratt_ho}
    Let $\b \in (\F_2)^{n-2}$ be as above and $\lb \in (\F_2)^{n\times n}$. The following are equivalent:
    \begin{itemize}
        \item $\lb \in H_\circ$;
        \item there exist $A \in \GL((\F_2)^2, +)$, $D \in \GL((\F_2)^d, +)$, and $B \in (\F_2)^{2 \times d}$ such that
        $$ 
        \lb = 
        \begin{pmatrix}
            A & B \\
            \mathbb{0}_{d,2} & D
        \end{pmatrix}
        $$
        and $\b D = \b$.
    \end{itemize} 
\end{theorem}


\section{Experiments on a 16-bit block cipher with 4-bit s-boxes}
\label{sec4}
As anticipated, the idea of this work is to design an SPN which is weak w.r.t.\ a differential attack based on an alternative parallel operation $\circ$ for which it is possible to show that the diffusion layer of the cipher belongs to $H_\circ$. We start by explaining explicitly what we mean by \emph{parallel}: letting 
$V=V_1\oplus...\oplus V_m$, with $V_i\simeq (\mathbb{F}_2)^n$ for  $i=1,...,m$, and $x\in V$, we can split $x$ in $m$ vectors $x_1,...,x_m$ of $n$ components each, and we can assume that the target SPN acting on a space of $m\times n$ bits contains $m$ s-boxes $S_1,\dots,S_m$ such that $S_1$ acts on $x_1$, $S_2$ on $x_2$ and so on. For this reason, we aim to mount an alternative differential attack using a sum $\circ$ acting as
\[
\begin{pmatrix}
    x_1 \\ \vdots \\ x_m
\end{pmatrix}
\circ
\begin{pmatrix}
    y_1 \\ \vdots \\ y_m
\end{pmatrix}
=
\begin{pmatrix}
    x_1 \circ_1 y_1 \\ \vdots \\ x_m \circ_m y_m
\end{pmatrix},
\]
where the sum $\circ_i$ acts on $V_i$.
As explained previously, the feasibily of the attack relies on an extension of Theorem \ref{teo:caratt_ho} to parallel sums.\\
In the absence of a general result in this sense, we have restricted our attention to the case where $W_{\circ_i}=\{k\mid k\in V_i \text{ is a weak key}\}$ has dimension $n-2$, for $i=1,...,m$, and we have performed some computational experiments using Magma~\cite{bosma1997magma}, which we describe below.
\subsection{The target cipher and its trapdoor}
Fixing $V = (\F_2)^{16}$, $n=4$ and $m=4$ and letting $\circ$ be the parallel sum defined applying each $4$-bit block the alternative operation $\circ_4$ defined by the vector $\b = (0, 1)$ (see Section~\ref{sec3}), we could check using Magma that the diffusion layer $\lambda$ defined in Fig.~\ref{figla} belongs to $H_\circ$, i.e., it is a permutation which is linear w.r.t.\ to both $+$ and $\circ$. Notice that the mentioned matrix, that will be chosen as the diffusion layer of the target SPN, is obtained from the cyclic shift of two $4\times4$ binary sub-matrices. For the benefit of the reader, we display the Cayley table of the $4$-bit operation $\circ_4$ induced by the vector $\b = (0, 1)$ in Fig.~\ref{figop}.
The entries in which $a\circ_4 b$ differs from $a+b$ are highlighted.

\begin{figure}
\[
\lb = 
\begin{pmatrix}
    0 & 0 & 0 & 0 & 0 & 0 & 0 & 0 & 0 & 0 & 0 & 0 & 1 & 0 & 0 & 0 \\
    0 & 0 & 1 & 0 & 0 & 0 & 1 & 0 & 0 & 0 & 1 & 0 & 0 & 1 & 1 & 1 \\
    0 & 0 & 1 & 0 & 0 & 0 & 1 & 0 & 0 & 0 & 1 & 0 & 0 & 0 & 0 & 0 \\
    0 & 0 & 0 & 0 & 0 & 0 & 0 & 0 & 0 & 0 & 0 & 0 & 0 & 0 & 0 & 1 \\
    1 & 0 & 0 & 0 & 0 & 0 & 0 & 0 & 0 & 0 & 0 & 0 & 0 & 0 & 0 & 0 \\
    0 & 1 & 1 & 1 & 0 & 0 & 1 & 0 & 0 & 0 & 1 & 0 & 0 & 0 & 1 & 0 \\
    0 & 0 & 0 & 0 & 0 & 0 & 1 & 0 & 0 & 0 & 1 & 0 & 0 & 0 & 1 & 0 \\
    0 & 0 & 0 & 1 & 0 & 0 & 0 & 0 & 0 & 0 & 0 & 0 & 0 & 0 & 0 & 0 \\
    0 & 0 & 0 & 0 & 1 & 0 & 0 & 0 & 0 & 0 & 0 & 0 & 0 & 0 & 0 & 0 \\
    0 & 0 & 1 & 0 & 0 & 1 & 1 & 1 & 0 & 0 & 1 & 0 & 0 & 0 & 1 & 0 \\
    0 & 0 & 1 & 0 & 0 & 0 & 0 & 0 & 0 & 0 & 1 & 0 & 0 & 0 & 1 & 0 \\
    0 & 0 & 0 & 0 & 0 & 0 & 0 & 1 & 0 & 0 & 0 & 0 & 0 & 0 & 0 & 0 \\
    0 & 0 & 0 & 0 & 0 & 0 & 0 & 0 & 1 & 0 & 0 & 0 & 0 & 0 & 0 & 0 \\
    0 & 0 & 1 & 0 & 0 & 0 & 1 & 0 & 0 & 1 & 1 & 1 & 0 & 0 & 1 & 0 \\
    0 & 0 & 1 & 0 & 0 & 0 & 1 & 0 & 0 & 0 & 0 & 0 & 0 & 0 & 1 & 0 \\
    0 & 0 & 0 & 0 & 0 & 0 & 0 & 0 & 0 & 0 & 0 & 1 & 0 & 0 & 0 & 0 \\
\end{pmatrix}
\]
\caption{The chosen diffusion layer}
\label{figla}
\end{figure}

\begin{figure}
\begin{center}
    \begin{tabular}{l|llllllllllllllll}
      $\circ_4$   & \hex{0} & \hex{1} & \hex{2} & \hex{3} & \hex{4} & \hex{5} & \hex{6} & \hex{7} & \hex{8} & \hex{9} & \hex{A} & \hex{B} & \hex{C} & \hex{D} & \hex{E} & \hex{F} \\ \hline
    \hex{0} & \normsum{\hex{0}} & \normsum{\hex{1}} & \normsum{\hex{2}} & \normsum{\hex{3}} & \normsum{\hex{4}} & \normsum{\hex{5}} & \normsum{\hex{6}} & \normsum{\hex{7}} & \normsum{\hex{8}} & \normsum{\hex{9}} & \normsum{\hex{A}} & \normsum{\hex{B}} & \normsum{\hex{C}} & \normsum{\hex{D}} & \normsum{\hex{E}} & \normsum{\hex{F}}\\ 
\hex{1} & \normsum{\hex{1}} & \normsum{\hex{0}} & \normsum{\hex{3}} & \normsum{\hex{2}} & \normsum{\hex{5}} & \normsum{\hex{4}} & \normsum{\hex{7}} & \normsum{\hex{6}} & \normsum{\hex{9}} & \normsum{\hex{8}} & \normsum{\hex{B}} & \normsum{\hex{A}} & \normsum{\hex{D}} & \normsum{\hex{C}} & \normsum{\hex{F}} & \normsum{\hex{E}}\\ 
\hex{2} & \normsum{\hex{2}} & \normsum{\hex{3}} & \normsum{\hex{0}} & \normsum{\hex{1}} & \normsum{\hex{6}} & \normsum{\hex{7}} & \normsum{\hex{4}} & \normsum{\hex{5}} & \normsum{\hex{A}} & \normsum{\hex{B}} & \normsum{\hex{8}} & \normsum{\hex{9}} & \normsum{\hex{E}} & \normsum{\hex{F}} & \normsum{\hex{C}} & \normsum{\hex{D}}\\ 
\hex{3} & \normsum{\hex{3}} & \normsum{\hex{2}} & \normsum{\hex{1}} & \normsum{\hex{0}} & \normsum{\hex{7}} & \normsum{\hex{6}} & \normsum{\hex{5}} & \normsum{\hex{4}} & \normsum{\hex{B}} & \normsum{\hex{A}} & \normsum{\hex{9}} & \normsum{\hex{8}} & \normsum{\hex{F}} & \normsum{\hex{E}} & \normsum{\hex{D}} & \normsum{\hex{C}}\\ 
\hex{4} & \normsum{\hex{4}} & \normsum{\hex{5}} & \normsum{\hex{6}} & \normsum{\hex{7}} & \normsum{\hex{0}} & \normsum{\hex{1}} & \normsum{\hex{2}} & \normsum{\hex{3}} & \emphsum{\hex{D}} & \emphsum{\hex{C}} & \emphsum{\hex{F}} & \emphsum{\hex{E}} & \emphsum{\hex{9}} & \emphsum{\hex{8}} & \emphsum{\hex{B}} & \emphsum{\hex{A}}\\ 
\hex{5} & \normsum{\hex{5}} & \normsum{\hex{4}} & \normsum{\hex{7}} & \normsum{\hex{6}} & \normsum{\hex{1}} & \normsum{\hex{0}} & \normsum{\hex{3}} & \normsum{\hex{2}} & \emphsum{\hex{C}} & \emphsum{\hex{D}} & \emphsum{\hex{E}} & \emphsum{\hex{F}} & \emphsum{\hex{8}} & \emphsum{\hex{9}} & \emphsum{\hex{A}} & \emphsum{\hex{B}}\\ 
\hex{6} & \normsum{\hex{6}} & \normsum{\hex{7}} & \normsum{\hex{4}} & \normsum{\hex{5}} & \normsum{\hex{2}} & \normsum{\hex{3}} & \normsum{\hex{0}} & \normsum{\hex{1}} & \emphsum{\hex{F}} & \emphsum{\hex{E}} & \emphsum{\hex{D}} & \emphsum{\hex{C}} & \emphsum{\hex{B}} & \emphsum{\hex{A}} & \emphsum{\hex{9}} & \emphsum{\hex{8}}\\ 
\hex{7} & \normsum{\hex{7}} & \normsum{\hex{6}} & \normsum{\hex{5}} & \normsum{\hex{4}} & \normsum{\hex{3}} & \normsum{\hex{2}} & \normsum{\hex{1}} & \normsum{\hex{0}} & \emphsum{\hex{E}} & \emphsum{\hex{F}} & \emphsum{\hex{C}} & \emphsum{\hex{D}} & \emphsum{\hex{A}} & \emphsum{\hex{B}} & \emphsum{\hex{8}} & \emphsum{\hex{9}}\\ 
\hex{8} & \normsum{\hex{8}} & \normsum{\hex{9}} & \normsum{\hex{A}} & \normsum{\hex{B}} & \emphsum{\hex{D}} & \emphsum{\hex{C}} & \emphsum{\hex{F}} & \emphsum{\hex{E}} & \normsum{\hex{0}} & \normsum{\hex{1}} & \normsum{\hex{2}} & \normsum{\hex{3}} & \emphsum{\hex{5}} & \emphsum{\hex{4}} & \emphsum{\hex{7}} & \emphsum{\hex{6}}\\ 
\hex{9} & \normsum{\hex{9}} & \normsum{\hex{8}} & \normsum{\hex{B}} & \normsum{\hex{A}} & \emphsum{\hex{C}} & \emphsum{\hex{D}} & \emphsum{\hex{E}} & \emphsum{\hex{F}} & \normsum{\hex{1}} & \normsum{\hex{0}} & \normsum{\hex{3}} & \normsum{\hex{2}} & \emphsum{\hex{4}} & \emphsum{\hex{5}} & \emphsum{\hex{6}} & \emphsum{\hex{7}}\\ 
\hex{A} & \normsum{\hex{A}} & \normsum{\hex{B}} & \normsum{\hex{8}} & \normsum{\hex{9}} & \emphsum{\hex{F}} & \emphsum{\hex{E}} & \emphsum{\hex{D}} & \emphsum{\hex{C}} & \normsum{\hex{2}} & \normsum{\hex{3}} & \normsum{\hex{0}} & \normsum{\hex{1}} & \emphsum{\hex{7}} & \emphsum{\hex{6}} & \emphsum{\hex{5}} & \emphsum{\hex{4}}\\ 
\hex{B} & \normsum{\hex{B}} & \normsum{\hex{A}} & \normsum{\hex{9}} & \normsum{\hex{8}} & \emphsum{\hex{E}} & \emphsum{\hex{F}} & \emphsum{\hex{C}} & \emphsum{\hex{D}} & \normsum{\hex{3}} & \normsum{\hex{2}} & \normsum{\hex{1}} & \normsum{\hex{0}} & \emphsum{\hex{6}} & \emphsum{\hex{7}} & \emphsum{\hex{4}} & \emphsum{\hex{5}}\\ 
\hex{C} & \normsum{\hex{C}} & \normsum{\hex{D}} & \normsum{\hex{E}} & \normsum{\hex{F}} & \emphsum{\hex{9}} & \emphsum{\hex{8}} & \emphsum{\hex{B}} & \emphsum{\hex{A}} & \emphsum{\hex{5}} & \emphsum{\hex{4}} & \emphsum{\hex{7}} & \emphsum{\hex{6}} & \normsum{\hex{0}} & \normsum{\hex{1}} & \normsum{\hex{2}} & \normsum{\hex{3}}\\ 
\hex{D} & \normsum{\hex{D}} & \normsum{\hex{C}} & \normsum{\hex{F}} & \normsum{\hex{E}} & \emphsum{\hex{8}} & \emphsum{\hex{9}} & \emphsum{\hex{A}} & \emphsum{\hex{B}} & \emphsum{\hex{4}} & \emphsum{\hex{5}} & \emphsum{\hex{6}} & \emphsum{\hex{7}} & \normsum{\hex{1}} & \normsum{\hex{0}} & \normsum{\hex{3}} & \normsum{\hex{2}}\\ 
\hex{E} & \normsum{\hex{E}} & \normsum{\hex{F}} & \normsum{\hex{C}} & \normsum{\hex{D}} & \emphsum{\hex{B}} & \emphsum{\hex{A}} & \emphsum{\hex{9}} & \emphsum{\hex{8}} & \emphsum{\hex{7}} & \emphsum{\hex{6}} & \emphsum{\hex{5}} & \emphsum{\hex{4}} & \normsum{\hex{2}} & \normsum{\hex{3}} & \normsum{\hex{0}} & \normsum{\hex{1}}\\ 
\hex{F} & \normsum{\hex{F}} & \normsum{\hex{E}} & \normsum{\hex{D}} & \normsum{\hex{C}} & \emphsum{\hex{A}} & \emphsum{\hex{B}} & \emphsum{\hex{8}} & \emphsum{\hex{9}} & \emphsum{\hex{6}} & \emphsum{\hex{7}} & \emphsum{\hex{4}} & \emphsum{\hex{5}} & \normsum{\hex{3}} & \normsum{\hex{2}} & \normsum{\hex{1}} & \normsum{\hex{0}}\\

    \end{tabular}
\end{center}
\caption{Cayley table of $\circ_4$}
\label{figop}
\end{figure}

The target cipher then features the 4-bit permutation $\gamma :(\F_2)^4 \rightarrow (\F_2)^4$ defined in Fig.~\ref{figsbox} as its s-box.
\begin{figure}
\begin{center}
    \begin{tabular}{|c|cccccccccccccccc|}
    \hline
        $x$      & \hex{0} & \hex{1} & \hex{2} & \hex{3} & \hex{4} & \hex{5} & \hex{6} & \hex{7} & \hex{8} & \hex{9} & \hex{A} & \hex{B} & \hex{C} & \hex{D} & \hex{E} & \hex{F} \\ \hline
        $x\gamma$ & \hex{0} & \hex{E} & \hex{B} & \hex{1} & \hex{7} & \hex{C} & \hex{9} & \hex{6} & \hex{D} & \hex{3} & \hex{4} & \hex{F} & \hex{2} & \hex{8} & \hex{A} & $5_x$\\
            \hline
    \end{tabular}
\end{center}
\caption{The chosen s-box $\gamma$}
\label{figsbox}
\end{figure}
Here each vector is interpreted as a binary number, most significant bit first. Precisely, four copies of $\gamma$ will act in a parallel way on the 16-bit block. Notice that the s-box $\gamma$ is \emph{optimal} in the sense of Leander and Poschmann~\cite{leander2007classification}. By computing the difference distribution table (DDT) of $\gamma$ with respect to XOR, we obtain the result displayed in Fig.~\ref{figDDT1}.
As it is known, $\gamma$ is differentially $4$-uniform, which is the best result for a permutation over $(\F_2)^4$ (see e.g.\ Leander and Poschmann~\cite{leander2007classification}).
\begin{figure}
\begin{center}
    \begin{tabular}{l|llllllllllllllll}
      +   & \hex{0} & \hex{1} & \hex{2} & \hex{3} & \hex{4} & \hex{5} & \hex{6} & \hex{7} & \hex{8} & \hex{9} & \hex{A} & \hex{B} & \hex{C} & \hex{D} & \hex{E} & \hex{F} \\ \hline
        \hex{0} &16 &$\cdot$ &$\cdot$ &$\cdot$ &$\cdot$ &$\cdot$ &$\cdot$ &$\cdot$ &$\cdot$ &$\cdot$ &$\cdot$ &$\cdot$ &$\cdot$ &$\cdot$ &$\cdot$ &$\cdot$\\

        \hex{1} &$\cdot$ &$\cdot$ &$\cdot$ &$\cdot$ &$\cdot$ &$\cdot$ &$\cdot$ &$\cdot$ &$\cdot$ &$\cdot$ &4 &4 &$\cdot$ &$\cdot$ &4 &4\\ 
        \hex{2} &$\cdot$ &$\cdot$ &$\cdot$ &$\cdot$ &$\cdot$ &$\cdot$ &$\cdot$ &$\cdot$ &2 &2 &2 &2 &2 &2 &2 &2\\ 
        \hex{3} &$\cdot$ &4 &4 &$\cdot$ &$\cdot$ &4 &$\cdot$ &4 &$\cdot$ &$\cdot$ &$\cdot$ &$\cdot$ &$\cdot$ &$\cdot$ &$\cdot$ &$\cdot$\\ 
        \hex{4} &$\cdot$ &$\cdot$ &4 &$\cdot$ &$\cdot$ &$\cdot$ &$\cdot$ &4 &$\cdot$ &$\cdot$ &2 &2 &$\cdot$ &$\cdot$ &2 &2\\ 
        \hex{5} &$\cdot$ &4 &$\cdot$ &$\cdot$ &$\cdot$ &4 &$\cdot$ &$\cdot$ &2 &2 &$\cdot$ &$\cdot$ &2 &2 &$\cdot$ &$\cdot$\\ 
        \hex{6} &$\cdot$ &$\cdot$ &$\cdot$ &$\cdot$ &$\cdot$ &$\cdot$ &4 &4 &2 &2 &$\cdot$ &$\cdot$ &2 &2 &$\cdot$ &$\cdot$\\ 
        \hex{7} &$\cdot$ &$\cdot$ &$\cdot$ &$\cdot$ &$\cdot$ &$\cdot$ &4 &4 &2 &2 &$\cdot$ &$\cdot$ &2 &2 &$\cdot$ &$\cdot$\\ 
        \hex{8} &$\cdot$ &$\cdot$ &$\cdot$ &4 &2 &2 &$\cdot$ &$\cdot$ &$\cdot$ &$\cdot$ &$\cdot$ &$\cdot$ &$\cdot$ &4 &2 &2\\ 
        \hex{9} &$\cdot$ &$\cdot$ &$\cdot$ &4 &2 &2 &$\cdot$ &$\cdot$ &$\cdot$ &$\cdot$ &$\cdot$ &$\cdot$ &4 &$\cdot$ &2 &2\\ 
        \hex{A} &$\cdot$ &2 &2 &$\cdot$ &2 &$\cdot$ &2 &$\cdot$ &$\cdot$ &2 &$\cdot$ &2 &$\cdot$ &2 &2 &$\cdot$\\ 
        \hex{B} &$\cdot$ &2 &2 &$\cdot$ &2 &$\cdot$ &2 &$\cdot$ &2 &$\cdot$ &2 &$\cdot$ &2 &$\cdot$ &$\cdot$ &2\\ 
        \hex{C} &$\cdot$ &2 &2 &$\cdot$ &2 &$\cdot$ &2 &$\cdot$ &$\cdot$ &2 &2 &$\cdot$ &$\cdot$ &2 &$\cdot$ &2\\ 
        \hex{D} &$\cdot$ &2 &2 &$\cdot$ &2 &$\cdot$ &2 &$\cdot$ &2 &$\cdot$ &$\cdot$ &2 &2 &$\cdot$ &2 &$\cdot$\\ 
        \hex{E} &$\cdot$ &$\cdot$ &$\cdot$ &4 &2 &2 &$\cdot$ &$\cdot$ &$\cdot$ &4 &2 &2 &$\cdot$ &$\cdot$ &$\cdot$ &$\cdot$\\ 
        \hex{F} &$\cdot$ &$\cdot$ &$\cdot$ &4 &2 &2 &$\cdot$ &$\cdot$ &4 &$\cdot$ &2 &2 &$\cdot$ &$\cdot$ &$\cdot$ &$\cdot$\\  
    \end{tabular}
\end{center}
\caption{DDT of $\gamma$ w.r.t.\ $+$}
\label{figDDT1}
\end{figure}
 However, if we compute the DDT using our new operation $\circ_4$ as  difference operator, we obtain
the result displayed in Fig.~\ref{figDDT2}.
\begin{figure}
\begin{center}
    \begin{tabular}{c|llllllllllllllll}
        $\circ_4$ & \hex{0} & \hex{1} & \hex{2} & \hex{3} & \hex{4} & \hex{5} & \hex{6} & \hex{7} & \hex{8} & \hex{9} & \hex{A} & \hex{B} & \hex{C} & \hex{D} & \hex{E} & \hex{F} \\ \hline
        \hex{0} &16 &$\cdot$ &$\cdot$ &$\cdot$ &$\cdot$ &$\cdot$ &$\cdot$ &$\cdot$ &$\cdot$ &$\cdot$ &$\cdot$ &$\cdot$ &$\cdot$ &$\cdot$ &$\cdot$ &$\cdot$\\ 
        \hex{1} &$\cdot$ &$\cdot$ &$\cdot$ &$\cdot$ &$\cdot$ &$\cdot$ &$\cdot$ &$\cdot$ &$\cdot$ &$\cdot$ &8 &$\cdot$ &$\cdot$ &$\cdot$ &8 &$\cdot$\\ 
        \hex{2} &$\cdot$ &$\cdot$ &$\cdot$ &$\cdot$ &$\cdot$ &$\cdot$ &$\cdot$ &$\cdot$ &4 &$\cdot$ &$\cdot$ &4 &4 &$\cdot$ &$\cdot$ &4\\ 
        \hex{3} &$\cdot$ &4 &4 &$\cdot$ &4 &$\cdot$ &$\cdot$ &4 &$\cdot$ &$\cdot$ &$\cdot$ &$\cdot$ &$\cdot$ &$\cdot$ &$\cdot$ &$\cdot$\\ 
        \hex{4} &$\cdot$ &4 &4 &$\cdot$ &4 &$\cdot$ &$\cdot$ &4 &$\cdot$ &$\cdot$ &$\cdot$ &$\cdot$ &$\cdot$ &$\cdot$ &$\cdot$ &$\cdot$\\ 
        \hex{5} &$\cdot$ &$\cdot$ &$\cdot$ &$\cdot$ &$\cdot$ &$\cdot$ &$\cdot$ &$\cdot$ &4 &$\cdot$ &$\cdot$ &4 &4 &$\cdot$ &$\cdot$ &4\\ 
        \hex{6} &$\cdot$ &$\cdot$ &$\cdot$ &$\cdot$ &$\cdot$ &$\cdot$ &$\cdot$ &$\cdot$ &$\cdot$ &8 &$\cdot$ &$\cdot$ &$\cdot$ &8 &$\cdot$ &$\cdot$\\ 
        \hex{7} &$\cdot$ &$\cdot$ &$\cdot$ &$\cdot$ &$\cdot$ &$\cdot$ &16 &$\cdot$ &$\cdot$ &$\cdot$ &$\cdot$ &$\cdot$ &$\cdot$ &$\cdot$ &$\cdot$ &$\cdot$\\ 
        \hex{8} &$\cdot$ &$\cdot$ &$\cdot$ &$\cdot$ &$\cdot$ &$\cdot$ &$\cdot$ &$\cdot$ &$\cdot$ &$\cdot$ &$\cdot$ &$\cdot$ &$\cdot$ &8 &8 &$\cdot$\\ 
        \hex{9} &$\cdot$ &$\cdot$ &$\cdot$ &8 &$\cdot$ &8 &$\cdot$ &$\cdot$ &$\cdot$ &$\cdot$ &$\cdot$ &$\cdot$ &$\cdot$ &$\cdot$ &$\cdot$ &$\cdot$\\ 
        \hex{A} &$\cdot$ &4 &4 &$\cdot$ &4 &$\cdot$ &$\cdot$ &4 &$\cdot$ &$\cdot$ &$\cdot$ &$\cdot$ &$\cdot$ &$\cdot$ &$\cdot$ &$\cdot$\\ 
        \hex{B} &$\cdot$ &$\cdot$ &$\cdot$ &$\cdot$ &$\cdot$ &$\cdot$ &$\cdot$ &$\cdot$ &4 &$\cdot$ &$\cdot$ &4 &4 &$\cdot$ &$\cdot$ &4\\ 
        \hex{C} &$\cdot$ &4 &4 &$\cdot$ &4 &$\cdot$ &$\cdot$ &4 &$\cdot$ &$\cdot$ &$\cdot$ &$\cdot$ &$\cdot$ &$\cdot$ &$\cdot$ &$\cdot$\\ 
        \hex{D} &$\cdot$ &$\cdot$ &$\cdot$ &$\cdot$ &$\cdot$ &$\cdot$ &$\cdot$ &$\cdot$ &4 &$\cdot$ &$\cdot$ &4 &4 &$\cdot$ &$\cdot$ &4\\ 
        \hex{E} &$\cdot$ &$\cdot$ &$\cdot$ &$\cdot$ &$\cdot$ &$\cdot$ &$\cdot$ &$\cdot$ &$\cdot$ &8 &8 &$\cdot$ &$\cdot$ &$\cdot$ &$\cdot$ &$\cdot$\\ 
        \hex{F} &$\cdot$ &$\cdot$ &$\cdot$ &8 &$\cdot$ &8 &$\cdot$ &$\cdot$ &$\cdot$ &$\cdot$ &$\cdot$ &$\cdot$ &$\cdot$ &$\cdot$ &$\cdot$ &$\cdot$\\ 
    \end{tabular}
\end{center}
\caption{DDT of $\gamma$ w.r.t.\ $\circ$}
\label{figDDT2}
\end{figure}
We can notice that $\gamma$ turns out to be differentially 16-uniform with respect to $\circ_4$; in particular, when the input difference is \hex{7}, the output difference becomes  \hex{6} with probability 1.  Beside this, it is clear from the table that the differential behaviour of the s-box is completely different when the alternative operation is considered and the map looks far away from being non-linear as necessary. \\
In our experiments described in the following section, we consider the SPN whose $i$-th round is obtained by the composition of the parallel application of the s-box $\gamma$ on every 4-bit block, of the diffusion layer $\lambda$ defined above, and of the XOR with the $i$-th round key.
\subsection{Brute-forcing differentials}
\begin{figure}[h!]
    \centering
    \includegraphics[scale=0.7]{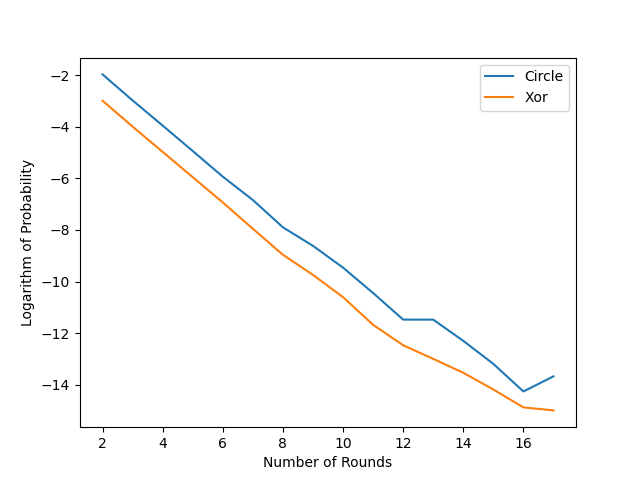}
    \caption{Best +-differential probability vs best $\circ$-differential probability}
    \label{fig}
\end{figure}

We study the difference propagation in the cipher in a long-key scenario, i.e., the key-schedule selects a random long key $k \in \mathbb F_2^{16r}$, where $r$ is the number of rounds. In order to mitigate the possible bias due to a particular key choice, we run our experiments taking the average over  $2^{15}$ random long-key generations.
This approach will provide us with a good estimate of the expected differential probability of the best differentials on this cipher. The experimental computations, carried out by \emph{brute-forcing} all the possible differentials, show that the best $i$-round differential for the classical XOR difference
is always less likely that the best $i$-round differential computed using the mentioned parallel operation, for $i=1, \ldots, 17$. The results, round per round, are displayed in Fig.~\ref{fig}. In particular, when $i=17$, the 
best $+$-differential is \hex{0060} $\rightarrow$ \hex{0700} with probability $2^{-14.993}$, while using the $\circ$ difference associated to $\b = (0, 1)$, the best 17-round $\circ$-differential is \hex{0070} $\rightarrow$\hex{0600} with probability $2^{-14.411}$.

Computational evidence shows  that similar results, even with a faster diffusion, can be obtained by choosing the diffusion layer of the cipher as a random matrix of $H_\circ$. This suggests that, in principle, every matrix of $H_\circ$ could represent a trapdoor diffusion layer for the cipher, with respect to a differential distinguishing attack which exploits the knowledge of the operation $\circ$.

\section{Open problems}\label{sec_fin}
In this paper, we have demonstrated that when the diffusion layer of an SPN exhibits linearity not only with respect to the XOR operation, as traditionally expected, but also in relation to an alternative operation stemming from a different translation group, this particular characteristic can be leveraged by a cryptanalyst to carry out a distinguishing attack employing alternative differentials. However, it can be quite challenging to discern which maps meet this criterion. What we require is an extension of Theorem~\ref{teo:caratt_ho} to the case of parallel operations, enabling the simultaneous targeting of all the s-boxes within the cipher, taking advantage of the lower non-linearity of the confusion layer. 
One potential approach to address this issue might involve attempting to represent the linear layer in a manner akin to the blocks demonstrated in Theorem \ref{teo:caratt_ho}. Based on our empirical findings, we offer the following hypothesis:

\begin{conjecture}
Let 
$V=V_1\oplus...\oplus V_m$, with $V_i\simeq (\mathbb{F}_2)^n$ for  $i=1,...,m$, and let $\circ=(\circ_1,...,\circ_m)$ be a parallel alternative operation as in Section \ref{sec4}, with $\dim(W_{\circ_i} )=n-2$ for all $i$. Then, the cardinality of $H_\circ$ is at least 
\[
m^3\cdot m!\cdot 3\cdot 2^{3n-6}\prod_{h=0}^{n-4}(2^{n-3}-2^h)[(m^2-m)2^{n^2-5n+6}-1].
\]
\end{conjecture}
This illustrates that $H_\circ$ may possess a sufficient size to contain matrices that appear to function as effective diffusion layers but, in reality, conceal trapdoor vulnerabilities.

Another crucial concern is the elimination of the assumption $d = n - 2$, as this would enable us to consider a broader range of operations. Nevertheless, as of the present writing, we are unaware of the existence of a more comprehensive version of Theorem~\ref{teo:caratt_ho} that eliminates the condition $d = n - 2$. Consequently, the prospect of an extension to the parallel case remains unknown.

In conclusion, it is evident that the influence of this approach on differential probabilities is intrinsically linked to the cipher's unique attributes. Computational evidence underscores the fact that even a minor modification in the design, such as altering the s-box or the diffusion layer, can have a profound influence on the resultant probabilities and outcomes. This heightened sensitivity to design specifics poses a challenge when attempting to establish general conjectures that can be universally applicable to different ciphers. Additionally, it is important to note that these resulting probabilities are heavily contingent on the fixed alternative operations. Notably, within $\mathbb{F}_2^{16}$, a vast number of approximately $2^{27}$ potential parallel alternative operations can be considered, working over 4-bit blocks.

A final thought to consider is the observation that, as we have demonstrated, alternative operations have the potential to diminish the resistance of an s-box to differential cryptanalysis. This is further exemplified by the fact that a 4-bit permutation, which is considered optimal (according to the criteria in Leander et al.~\cite{leander2007classification}), exhibits the lowest possible differential uniformity when coupled with the operation defined in Fig. \ref{figop}. This raises an interesting open problem: the complete analysis of differential properties concerning alternative operations of various s-boxes, akin to what has been explored for the 4-bit permutations with respect to modular addition~\cite{zajac2020cryptographic}. Even when focusing on small dimensions like 4-bit permutations, this undertaking requires some efforts. It is important to note that in this context, there are 106 possible operations available (as detailed in Calderini et al.~\cite[Table 1]{calderini}), including the XOR. Moreover, within the same affine-equivalence class of a given s-box, different functions may exhibit varying behavior with respect to a fixed alternative operation.


We believe that the experimental results of this paper show why the mentioned problems can be of interest in this area of research in cryptanalysis.

\section*{Acknowledgements}
This work has been accepted for presentation at CIFRIS23, the Congress of the Italian association of cryptography ``De Componendis Cifris". M.\ Calderini and R.\ Civino are members of INdAM-GNSAGA
 (Italy).  
 
 \section*{Funding information} 
 R.\ Civino is funded by the Centre of excellence
 ExEMERGE at University of L'Aquila.

 \section*{Conflict of interest}
 Authors state no conflict of interest.

\bibliographystyle{vancouver}
\bibliography{refs}

\end{document}